\documentclass[a4paper,12pt]{article} 
\usepackage{times}
\usepackage[pdftex]{graphicx}
\usepackage{common}
\usepackage{url}
\usepackage{hyperref}
\usepackage{multirow}
\usepackage{array}
\usepackage{natbib}
\usepackage{subfig}
\bibpunct{(}{)}{;}{a}{,}{,}

\title{Using \emph{tf-idf} as an edge weighting scheme in user-object bipartite networks}

\author{
Sorin Alupoaie, P\'{a}draig Cunningham \\
School of Computer Science \& Informatics\\
University College Dublin\\
\small{\texttt{sorin.alupoaie@ucdconnect.ie,padraig.cunningham@ucd.ie}} \\
}

\begin{document}
\maketitle

\begin{abstract}
Bipartite user-object networks are becoming increasingly popular in representing
user interaction data in a web or e-commerce environment. They have certain
characteristics and challenges that differentiates them from other bipartite
networks. This paper analyzes the properties of five real world user-object
networks. In all cases we found a heavy tail object degree distribution with
popular objects connecting together a large part of the users causing
significant edge inflation in the projected users network. We propose a novel
edge weighting strategy based on \emph{tf-idf} and show that the new scheme
improves both the density and the quality of the community structure in the
projections. The improvement is also noticed when comparing to partially random
networks.
\end{abstract}


\section{Introduction}

Social networks are becoming increasingly popular for modeling interactions
between nodes or individuals of the same kind (one-mode or unipartite) and there
is a large amount of work recently focusing on various aspects of social
networks. However some real-world networks have a more heterogeneous structure
where two kinds of individuals (or items) interact with each other with ties
formed between individuals of different types only. Bipartite networks
(two-mode) are a natural fit for these sort of systems as they are represented
as a graph with two disjoint sets of nodes and edges exists only between nodes
from different sets. There are many examples of complex networks that have a
bipartite structure, such as actor-movie networks, where actors are linked to
movies they played in \citep{guillaume2004bipartite}, authoring or collaboration
networks, where authors are linked to the paper they published
\citep{newman2001scientific1}, human sexual networks, consisting of men an women
\citep{liljeros2001web} and the metabolic networks between chemical reactions
and metabolites \citep{jeong2000large}. Even though there is a lot more focus on
social networks in the scientific community, research on bipartite or two-mode
networks has moved forward in the last couple of years. 

User-object networks are emerging as a special class of bipartite networks with
certain characteristics that are different than the relatively well studied
author-paper and actor-movie networks. Users interact continuously with objects,
based on their own selection and preference. This representation is appropriate
for modeling activities in web and e-commerce environments. For example the
social tagging, music listening activity or movie watching can be modelled as an
user-object network. In certain cases these networks can be also referred as
consumer-product networks where an edge links a consumer with a product when the
former buys or views that particular product \citep{huang2007analyzing}. 

One particular characteristic of user-object networks is the presence of a
significant number of highly popular objects reflected by the heavy tail
distribution of the objects degree \citep{shang10empirical}. The heavy tail is
believed to be formed through preferential attachment
\citep{barabasi2002evolution}, where users tend to interact more with objects
that are already popular. These popular objects have a significant effect on the
properties of both bipartite and projected networks, causing significant link
inflation in the latter. On the other side they are a poor indicator of users
interests, while unpopular objects are the best at describing common tastes
shared by users \citep{shang10empirical}. In this paper we are proposing a
method of (re)assigning weights to edges in a bipartite user-object network
based on the popular \emph{tf-idf} method that will reduce the effect of popular
objects while taking into account user's preferences. We validate the new method
by showing that both density and community structure of the projected users
network is improved compared to both the original projected network and the
network generated partially in a random fashion.

The remainder of the paper is structured as follows. Next section reviews the
state of the art literature on bipartite and user-object networks. In section 3
we provide a comprehensive analysis of user-object networks and their properties
and challenges, including density and clustering. Section 4 proposes the new
weighting scheme based on \emph{tf-idf}. In section 5 we validate the proposed
approach by comparing densities and modularities of the projected users network
to the random case. Finally, we conclude the paper in section 6 by summarizing
our findings and pointing out future research directions.


\section{Related work}

One popular approach when analyzing two-mode networks is to transform them into
one-mode networks through a method known as projection and then use the methods
available for social networks. \citet{newman2001scientific1} and
\citet{barabasi2002evolution} study scientific collaboration networks between
authors of scientific papers by projecting the authoring network (author-paper)
to an unweighted co-authoring network (author-author). A tie is defined between
two authors if they have at least one collaboration together, without taking
into account the frequency of collaboration between the authors. In order to
avoid the loss of information through unweighted projection, some authors
proposed a weighting method. \citet{ramasco2006social} and
\citet{li2007evolving} use a simple mechanism to assign weights, based on the
number of times those authors collaborated. Other papers propose a different
approach to weighting. In \citet{newman2001scientific2} the contribution of a
co-authored article to the weight of an edge between two authors depends on the
degree of the article. This is based on the assumption that a low degree article
defines a stronger relationship between authors than a high degree one.
\citet{li2005weighted} considers a saturating effect where the increase in
strength of a relationship between two authors slows down when more articles
are added, as they already know each other well after writing some papers
together.

In order to avoid loss of information through projection (weighted or
unweighted), several features and methods have been proposed for bipartite
networks, most of them borrowed from social networks.
\citet{borgatti1997network} studies visualization techniques of bipartite
networks and defines several properties for these type of networks like density
and centrality. \citet{latapy2008basic} uses several statistics to analyze
two-mode data and proposes new bipartite properties like clustering coefficient
and redundancy coefficient. \citet{lind2005cycles} and
\citet{zhang2008clustering} redefine the clustering coefficient for bipartite
networks by considering squares (4-cycles) instead of triangles, which are not
possible in a two-mode configuration. \citet{opsahl2011triadic} proposes a new
measure of clustering based on the notion of triadic closure defined between
three nodes of the same type. Several community detection methods have been
proposed for bipartite networks \citep{fortunato2010community}. However most of
these methods are not optimized for large networks on the scale seen in web or
e-commerce environment. Therefore when detecting communities in bipartite
networks a widely used approach is to project them on one of the node sets and
apply scalable community detection methods from social networks on the
projection.

The research work mentioned above focused either on specific networks like
collaboration network or generally on bipartite networks. They didn't consider
user-object networks as a particular class of bipartite networks. On the other
side, especially in a web or online environment user-object networks with their
particularities are becoming increasingly popular in modeling the interaction
between users (e.g. customers, listeners, watchers etc) and the online system
(e.g. e-commerce, music, video etc). \citet{huang2007analyzing} is analyzing the
bipartite consumer-product graphs representing sales transactions in an
e-commerce setting. They found a larger than expected average path length and
tendency of customers to cluster according to their purchases. A new
recommendation algorithm is proposed based on these findings.
\citet{grujic2009mixing} uses a bipartite representation of interactions between
users and web databases to study patterns of clustering based on users common
interests. They found a power law degree distribution of objects and a
disassortative mixing pattern with high degree (active) users interacting mostly
with low degree (unpopular) objects and low degree (inactive) users interacting
mostly with high degree objects (popular). The authors applied a spectral
clustering method on the weighted projected network and found communities
relevant to subjects of common interests.

Recently, \citet{shang10empirical} did an empirical analysis of two web-based
user-object networks collected from two large-scale web sites and found the same
power law degree distribution for objects and disassortative mixing pattern. A
new property is proposed called collaborative similarity capturing the diversity
of tastes based on the collaborative selection. For the lower-degree objects the
authors found a negative correlation between object collaborative similarity
(how similar are the users interacting with a specific object) and the object
degree. Therefore unpopular (low degree) objects are considered a good indicator
for the users common interests, while popular objects are less relevant.
Starting from this observation we propose a new weighting scheme for user-object
bipartite networks that will increase the relevance of unpopular object in the
network and decrease the importance of popular ones.


\section{Characteristics of user-object networks} \label{sec:ubnetworks}

A user-object network can be represented as a graph $G = (U, O, E)$, where $U$
is the users set, $O$ is the objects set and $E \subseteq (U \times O)$ is the
set of edges between users and objects. We denote by $n_u = |U|$ and $n_o = |O|$
the number of users and the number of objects, respectively. We denote by $m =
|E|$ the number of links or edges in the bipartite network. We define $\langle
k_U \rangle$ ($\langle k_O \rangle$) as the average degree of users (objects)
and the density of the network as $\delta(G) = \frac{m}{n_o n_u}$.

In user object networks interactions between the two types of nodes, $U$ and
$O$, are event-driven, occurring continuously, and often the same edge is
reinforced multiple times. This reinforcement can be represented as a weight
assigned to the edge between the two interacting nodes. On the other side, in
most bipartite networks there is only one interaction between two nodes (an
actor can only play once in a movie).

The behaviour of nodes is also particular. Users are active while objects are
passive. This is similar to author-paper networks but different than other
bipartite networks such as human sexual networks where both nodes are active
\citep{shang10empirical}.

\begin{figure}[t]
\centering
  \includegraphics[width=8cm]{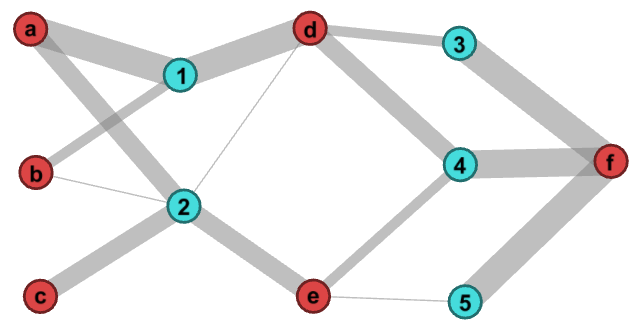}  
  \caption{Example of weighted bipartite network where users (blue nodes) are
  connected with objects (red nodes).
  }
\label{fig:example}
\end{figure}

Figure \ref{fig:example} illustrates a small example of a weighted bipartite
network where five users (blue nodes) are interacting with six objects (red
nodes). Thickness of an edge shows how strong the interaction is between the
user and the object (i.e. how often the user `touches` the object). The degree
of an user $u \in U$, denoted by $d(u)$, represents the number of objects
connected to that user. Similarly, the degree of an object $o \in O$, denoted by
$d(o)$, is the number of users interacting with that object. In our example,
user $1$ interacts with three objects $a$, $b$ and $d$ (has a degree of three).
Out of these, objects $a$ and $b$ are local as they are only connected to two
users ($1$ and $2$), while object $d$ reaches most of the network by being
connected to four out of a total of five users. Therefore objects $a$ and $b$
are more relevant to particular interests of user $1$ than object $d$.

\subsection{Data}

We are using five real-world data sets and look at the characteristics of the
related user-object networks such as basic statistics, degree distributions and
clustering. The first one is a subset of the Last.fm Million Song Dataset
\citep{Bertin-Mahieux2011}, where users are connected to tags, with an edge
defined between an user and a tag if that user assigned the tag to an artist.

The second dataset was extracted from Twitter and includes tweets related to US
Politics, covering the period 01/01/2012 - 19/01/2012. This is a subset of a
bigger dataset including the US Presidential Campaign and Elections during
2012. The users in the data are the Presidential candidates, governors,
senators, political organisations, and journalists. The objects are the web
domains included in their tweets. An edge is formed between an user and a domain
if the domain was tweeted by that particular user.

The third dataset is a subset of a bigger database published by
Audioscrobbler.com containing information about users and the music they
listened. In the network representation, an user is connected to an artist
with edges weighted based on how often they listen songs from that artist. The
Movielens dataset includes ratings (from 1 to 5) of movies by 2000 users,
represented as a network. For the purpose of this paper only edges with ratings 4 and 5
were considered (movies that users actually liked). Finally, the last dataset
comes from the popular social bookmarking web site delicious.com and includes
the network of 973 users and the tags they used for bookmarking.

\begin{table}[!h] \centering
\begin{tabular}{ l | c c c c c c }
 Data & $n_u$ & $n_o$ & $m$ & $\langle k_U \rangle$ & $\langle k_O \rangle$ &
 $\delta(G)$ \\
\hline
Last.fm & 1,892 & 9,748 & 35,813 & 18 & 3 & 0.19\% \\
Twitter & 1,842 & 3,744 & 13,864 & 7 & 3 & 0.2\% \\
Audioscrobbler & 183 & 21,443 & 39,195 & 214 & 1 & 1\% \\
Movielens & 2,000 & 3,336 & 192,922 & 96 & 57 & 2.9\% \\
Delicious & 973 & 28,695 & 126,007 & 129 & 4 & 0.45\% \\
\end{tabular}
\caption{Statistics of the five real-world datasets.}
\label{tab:datasets}
\end{table}

The basic statistics for these networks are presented in Table
\ref{tab:datasets}. All networks have relatively low density (they are sparse),
between 0.19\% and 2.9\%. This is important because it allows for optimization
of clustering and community detection algorithms. As we will see below this is
not the case for the projections of these bipartite networks.

\subsection{Projection}

Due to the shortage in tools and notions available for two-mode networks a
common approach is to transform such a network into its projection by using one
of its sets of nodes as a base. For example one can project the network on the
users $U$ (objects $O$) by setting a (weighted) edge between two users (objects)
that have at least one object (user) in common. Based on the number of common
objects (users), weights can be assigned to edges in the projected network. We
define the users-projected network as $G_U = (U, E_U)$ and objects-projected
network as $G_O = (O, E_O)$. The number of edges is $m_U = |E_U|$ and $m_O =
|E_O|$, respectively. The density of the projected networks is $\delta(G_U) =
\frac{2 m_U}{n_u (n_u-1)}$ and $\delta(G_O) = \frac{2 m_O}{n_o (n_o-1)}$,
respectively.

One of the unwanted effects of projection is an inflation of the number of links
in the resulted network \citep{latapy2008basic}, which can be noticed in random
networks as well. \citet{newman2001random} have shown that projecting a random
bipartite graph can result in very dense networks with high clustering
coefficient. In the real world case these dense networks are limiting the
computations that could be performed to extract properties and detect community
structure. 

\begin{table}[!h] \centering
\begin{tabular}{ l | c c c c }
Data & $m_U$ & $m_O$ & $\delta(G_U)$ & $\delta(G_O)$ \\
\hline
Last.fm & 686,536 & 322,226 & 38.3\% & 0.6\% \\
Twitter & 446,892 & 88,044 & 26.3\% & 1.2\% \\
Audioscrobbler & 10,453 & 11,253,485 & 62.7\% & 4.9\% \\
Movielens & 1,786,647 & 2,879,932 & 89.3\% & 51.7\% \\
Delicious & 395,835 & 7,378,472 & 83.7\% & 1.8\% \\
\end{tabular}
\caption{Statistics of projections of the real-world bipartite networks.}
\label{tab:datasets_prj}
\end{table}

For example, as shown in Table \ref{tab:datasets_prj}, the density is increasing
when projecting the real world networks on both user-nodes and object-nodes.
However, the inflation of links is considerably (up to 200 times) larger for the
projection on user-nodes, while the projection on object-nodes has a density in
most cases only several times higher than the original bipartite network. As we
will see below this is caused by the small number of very popular objects, the
tail of the power law distribution of object degrees.

\subsection{Degree distributions}

\begin{figure}[t] \centering
  \begin{tabular}{cc} 
  \subfloat[Last.fm
  Users]{\label{fig:degrees_lastfm_users}\includegraphics[width=5cm]{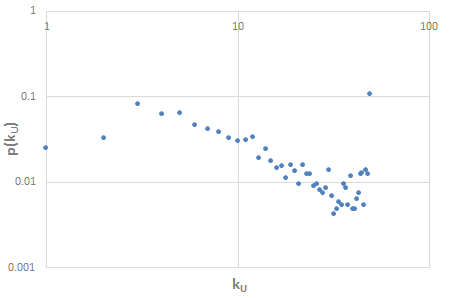}}
  &
  \subfloat[Twitter
  Users]{\label{fig:degrees_twitter_users}\includegraphics[width=5cm]{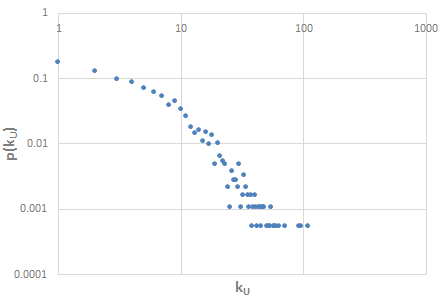}} 
  \\
  \subfloat[Audioscrobbler
  Users]{\label{fig:degrees_audioscrobbler_users}\includegraphics[width=5cm]{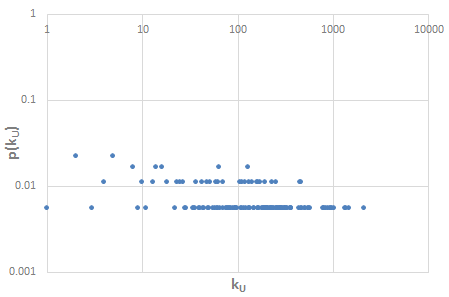}}
  &
  \subfloat[Movielens
  Users]{\label{fig:degrees_movielens_users}\includegraphics[width=5cm]{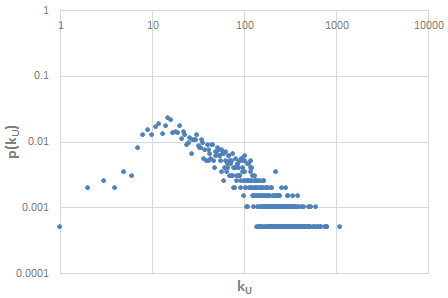}}
  \\
  \end{tabular}
  \begin{tabular}{c} 
  \subfloat[Delicious
  Users]{\label{fig:degrees_delicious_users}\includegraphics[width=5cm]{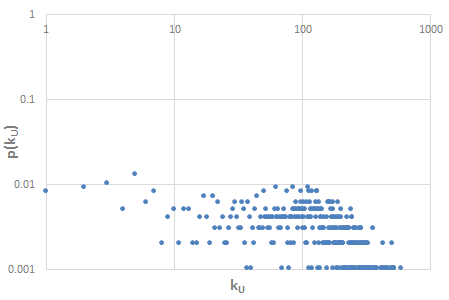}}
  \\ 
  \end{tabular}
  \caption{Degree distributions of users for the real world data sets}
  \label{fig:degrees_users}
\end{figure}

One of the most important structural property of networks is the degree
distribution. In the case of bipartite networks and users-objects networks
specifically, distributions of both types of nodes are considered.
According to previous studies, users degree generally follow an exponential
distribution
\citep{latapy2008basic}, or stretched exponential \citep{laherrere1998stretched}
at most, as shown in \citet{shang10empirical} or \citet{grujic2009mixing}. On
the other side, according to the same papers, the degree distribution of objects
follow a heavy tail distribution (power law or similar). This has important
consequences on the structure of the projected network and also on the process
of clustering the user nodes. Other bipartite networks like author-paper or
actor-movie doesn't exhibit this property as there is an inherent constraint on
how many authors can write a paper or how many actors can play in a movie.

\begin{figure}[t] \centering
  \begin{tabular}{cc} 
  \subfloat[Last.fm
  Tags]{\label{fig:degrees_lastfm_objects}\includegraphics[width=5cm]{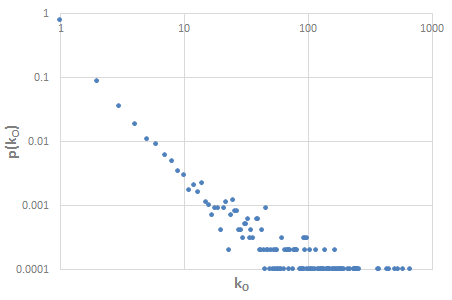}}
  & 
  \subfloat[Twitter
  Domains]{\label{fig:degrees_twitter_objects}\includegraphics[width=5cm]{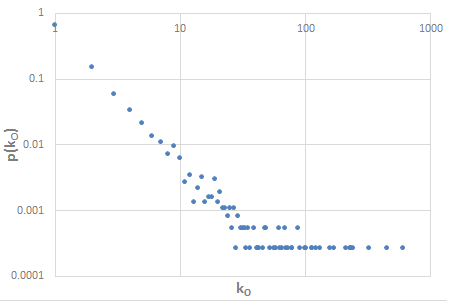}}
  \\
  \subfloat[Audioscrobbler
  Artists]{\label{fig:degrees_audioscrobbler_objects}\includegraphics[width=5cm]{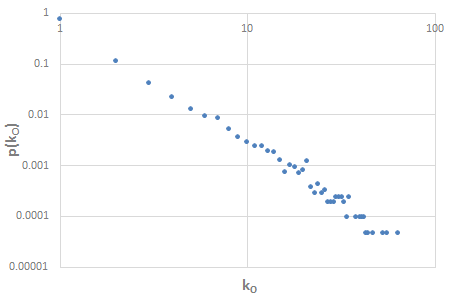}}
  &
  \subfloat[Movielens
  Movies]{\label{fig:degrees_movielens_objects}\includegraphics[width=5cm]{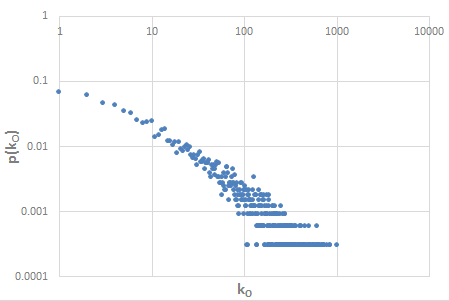}}
  \\
  \end{tabular}
  \begin{tabular}{c} 
  \subfloat[Delicious
  Tags]{\label{fig:degrees_delicious_objects}\includegraphics[width=5cm]{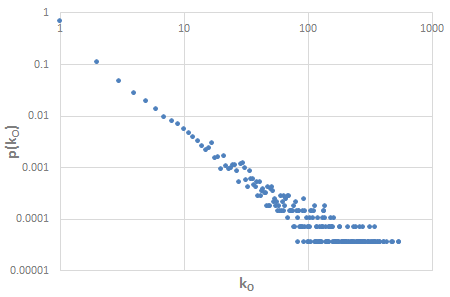}}
  \\
  \end{tabular}
  \caption{Degree distributions of objects for the real world data sets}
  \label{fig:degrees_objects}
\end{figure}

Degree distributions of our sample user-object networks are shown in figures
\ref{fig:degrees_users} and \ref{fig:degrees_objects}. In order to find the
closest model that fits the data, for each such empirical distribution we
evaluate the goodness of fit of multiple distribution models by using
loglikelihood ratios as described in \citet{clauset2009power}. The best fit is
found through comparing multiple models two by two: exponential, power law,
log-normal and stretched exponential. In each comparison we calculate the
loglikelihood ratio between two candidate distributions. This value will
indicate which model is a better fit for the data. Also a p-value is calculated
for each comparison showing how significant the result is.

\begin{table}[!h] \centering
\begin{tabular}{ l | c c | c c }
\multirow{2}{*}{Data} & \multicolumn{2}{c}{Users} & \multicolumn{2}{c}{Objects}
\\
\cline{2-5}
& Best fit & Parameters & Best fit & Parameters \\
\hline
Last.fm & exponential & $\lambda = 0.054$ & lognormal & $\mu = -33.18$ \\
Twitter & lognormal & $\mu = 1.41$ & powerlaw & $\alpha = 2.05$ \\
Audioscr & exponential & $\lambda = 0.0046$ & lognormal & $\mu = -2.75$ \\
Movielens & lognormal & $\mu = 4.07$ & lognormal & $\mu = 2.84$ \\
Delicious & exponential & $\lambda = 0.0081$ & lognormal & $\mu = -9.34$ \\
\hline
\end{tabular}
\caption{Best fit for degree distributions, users and objects.}
\label{tab:dist_bestfit}
\end{table}

In table \ref{tab:dist_bestfit} we show the best distribution fit and the
related parameters for each dataset. While degree distribution for users (active
nodes) don't expose a heavy tail in most cases (exponential), the degree
distribution for objects (passive nodes) fits to one of the distribution models
with a tail that is not exponentially bounded (power law or lognormal).
Due to the number of high degree values (highly popular objects), a heavy tail
with heterogeneous distribution of degrees has a significant impact on the
structure of the projected network and consequently on the community structure.

\subsection{Clustering and impact of highly popular objects}

Clustering or community structure \citep{newman2004finding} is another property
of user-object or consumer-product networks. Users or consumers tend to cluster
together based on the common interests represented by the common objects they
are interacting with \citep{huang2007analyzing}. This property sits at the core
of the recommender systems based on collaborative filtering
\citep{breese1998empirical}. However, detecting communities in bipartite
networks is somehow challenging with few optimized options available at large
scale \citep{fortunato2010community}.

A popular approach is to project the bipartite network on one of its node sets
($U$ or $O$) and apply the community detection algorithm on the projected
network. As seen above, the projection of user-object networks where degrees of
objects follow a power law distribution is causing edge inflation in the
resulted network by increasing significantly the density of edges (hundreds of
times more). This affects the clustering in two ways. First, it `dilutes` the
clusters and decreases the quality of the community structure because of a
highly homogeneous distribution of edges among the nodes
\citep{fortunato2010community}. Second, it causes a drop in performance of
community detection algorithms, most of them optimized for sparse networks. For
example, popular algorithms like \citet{clauset2004finding},
\citet{raghavan2007near} or \citet{blondel2008fast} scale extremely well on
sparse data but are much less efficient when networks are very dense (number of
links $m$ are much larger than number of nodes $n$).

In our case, due to the heavy-tail distributions of objects, user-projections of
all five real-world networks are extremely dense ranging from 26\% to 89\%
(Table \ref{tab:datasets_prj}) causing difficulties such as those described
above. For example, by looking at the most popular objects in Table
\ref{tab:topdegrees} one can see that the highest degree node in each case
connects together more than a third of users and in some cases even 50\%.

\begin{table}[!h] \centering
\begin{tabular}{cc}
 
\subfloat[Last.fm user-tag] {
\begin{tabular}{ l | c p{1.2cm}}
Tag & Degree & Users linked\\
\hline
rock & 673 & 35.5\% \\
pop & 585 & 30.9\% \\
alternative & 532 & 28.1\% \\
\end{tabular}
}

\subfloat[Twitter user-domain] {
\begin{tabular}{ l | c p{1.2cm} }
Domain & Degree & Users linked \\
\hline
twitter.com & 611 & 33.2\% \\
facebook.com & 451 & 24.5\% \\
youtube.com & 324 & 17.6\% \\
\end{tabular}
}
\end{tabular}

\begin{tabular}{cc}
\subfloat[Audioscrobbler user-artist] {
\begin{tabular}{ l | c p{1.2cm} }
Artist & Degree & Users linked \\
\hline
Radiohead & 65 & 35.5\% \\
Coldplay & 56 & 30.6\% \\
The Beatles & 53 & 28.9\% \\
\end{tabular}
}

\subfloat[Movielens user-movie] {
\begin{tabular}{ l | c p{1.2cm} }
Movie & Degree & Users linked \\
\hline
American Beauty & 1009 & 50.4\% \\
Star Wars IV & 855 & 42.7\% \\
Star Wars V & 841 & 42\% \\
\end{tabular}
}
\end{tabular}

\begin{tabular}{c}
\subfloat[Delicious user-tag] {
\begin{tabular}{ l | c p{1.2cm} }
Tag & Degree & Users linked \\
\hline
design & 546 & 56.1\% \\
video & 543 & 55.8\% \\
google & 489 & 50.2\% \\
\end{tabular}
}

\end{tabular}
\caption{Top objects by degree with percentage of users each object connects.}
\label{tab:topdegrees}
\end{table}

On the other side, these highly popular objects contain very little or very high
level information about the particularities of the adjacent users and are not
very meaningful for grouping users together. 

For example, while Last.fm tags like 'rock', 'pop' or 'alternative' might
contain some high level information on the users groups, it doesn't bring
specific information about these. In the meantime, unpopular (degree 5) tags
like 'symphonic rock', 'david bowie' or 'true norwegian black metal' are showing
narrow user interests. Also, highly popular tweeted domains like 'twitter.com'
or 'facebook.com' are very broad with little clustering information, while
unpopular domains like 'radioamerica.org' or 'seacoastonline.com' are much more
specific. This is in line with \citet{shang10empirical} who found that users
connected to unpopular objects have much higher similarity to each other than
the average. Therefore unpopular objects are considered a better indicator for
users common interests than popular ones.

One simple way to handle this problem is to remove the objects with the highest
degree, but this will result in loss of information overall, especially in cases
when these objects are reinforced frequently by some users. The proposed method
in section \ref{sec:tfidf} will address this issue by trying to find a balance
between how often an user interacts with an object and how popular an object is
in the network.


\section{Edge weighting using \emph{tf-idf}} \label{sec:tfidf}

One of the most common methods used in information retrieval is the \emph{term
frequency -– inverse document frequency} (or \emph{tf--idf}), a weighing scheme
for quantifying the importance of a term to a document in a collection
\citep{salton1988term}. The tf--idf weight is assigned to each term-document
pair and is proportional to the relative frequency of the word in the document,
adjusted by the proportion of that word over the entire collection. This method
and variations of it are used by search engines as a central component for
ranking the relevance of a document to a query, but also in other applications
like text summarization and classification.

The tf--idf weight of a term $t$ for a document $d$ in a collection $D$ is
defined as the product of the term frequency and inverse document frequency:

\begin{equation}
\label{formula:tfidf}
w_d(t) = f_d(t) \times log(\frac{\mid D\mid}{|\{d \in D : t \in d\}|})
\end{equation}

The term frequency $f_d(t)$ in equation \ref{formula:tfidf} can be the raw or
relative frequency of the term $t$ in the document $d$. The inverse document
frequency (second part of the product) measures whether the term is common or
rare across all documents in the collection $D$ and is calculated by taking the
logarithm of the fraction of the size of the collection $D$ to the number of
documents containing the term $t$.

We propose a similar approach for (re)weighting a user-object bipartite network
in order to reduce the impact of highly popular objects on projection and
clustering. Let us replace the documents in the tf--idf scheme with users and
terms with objects, and define a weighting scheme as below:

\begin{equation}
\label{formula:weight}
w_{new}(u,o) = f(u,o) \times
log(\frac{|U|}{d(o)})
\end{equation}

The new weight (\ref{formula:weight}) of an edge between an user and an object
can be recalculated as the product of the normalized object frequency and the
logarithm of the inverse object frequency across all users. The former
represents the number of object-user interactions (edge weight) with some
normalization technique applied. Normalization is required to prevent a bias
towards more active users. For example one can normalize the weight of an edge
between an user and an object by taking the ratio of its value to the maximum
weight of any given edge connected to that user (\ref{formula:tfmax}).

\begin{equation}
\label{formula:tfmax}
f(u,o) = \frac{w(u,o)}{max\{w(u,p) : p \in N(u)\}}
\end{equation}

The last part of the product in equation \ref{formula:weight} is similar to the
concept of \emph{inverse document frequency} and measures whether the object is
common or rare across all users. It is calculated by dividing the total number
of users by the (unweighted) degree of the object (number of users interacting
with the object), and taking the logarithm of this value.

The re-weighting scheme presented above assigns higher values to frequent
interactions between users and objects offset by how popular an object is within
the collection of users. Unpopular objects increases the weight of an
interaction while very popular ones make the interaction less significant.

Once the new weights have been calculated, thresholding can be applied to filter
out irrelevant edges. As we'll see in section \ref{sec:eval}, this reduces
significantly the edge inflation in the projected networks while increasing the
quality of partitions resulted from community detection and improving the
performance of the algorithms.

\begin{figure}[t] \centering
  \begin{tabular}{cc} 
    \subfloat[Example network with tf-idf
  weights]{\label{fig:example_weighted_tfidf}\includegraphics[width=6cm]{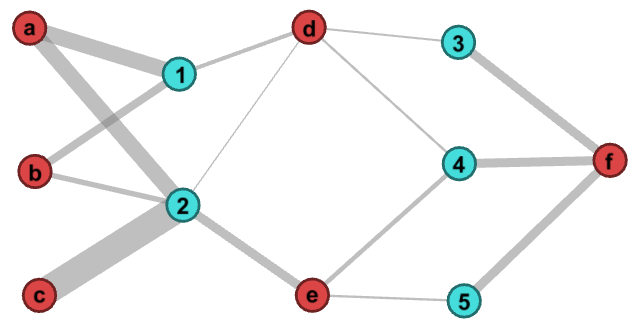}}
  & \subfloat[Filtered network $\tau =
  0.3$]{\label{fig:example_filtered}\includegraphics[width=6cm]{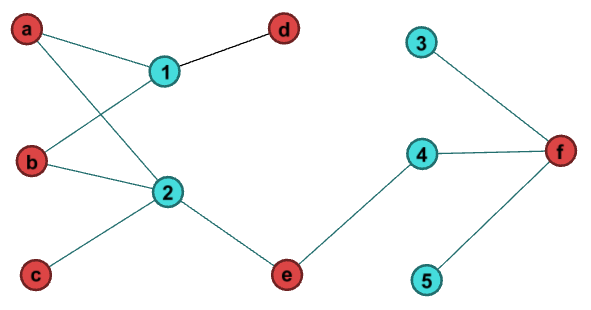}}
  \end{tabular}
  \caption{Example of user-object network with \emph{tf-idf} weighting and
  filtering applied}
  \label{fig:example_all}
\end{figure}

Figure \ref{fig:example_all} illustrates the network example from section
\ref{sec:ubnetworks} (figure \ref{fig:example}) with the proposed \emph{tf-idf}
method applied. Weights are assigned to edges based on their initial value and
transformed through equation \ref{formula:weight}. In our example they vary
between 0.1 and 2.3. The thickness of an edge is proportional to the resulted
weight after \emph{tfidf} is applied. As illustrated in figure
\ref{fig:example_weighted_tfidf}, the lowest values are assigned to edges
pointing at the highest degree object $d$. This object is connected to four out
of five users. The highest weights are assigned to edges connecting users to
unpopular or local objects, for example $a$ and $b$. The initial weight is
also taken into consideration with higher weights assigned to edges having
higher initial value. In our example, edge $1d$ has higher \emph{tfidf} weight
than $2d$, even though both edges end in the same node $d$.

After filtering out all edges with weights smaller than a threshold (in this
case $\tau = 0.3$), the resulted network has a simpler and more intuitive
structure, containing only relevant edges (figure \ref{fig:example_filtered}).
In the next section, we will validate the proposed approach and confirm this
observation on large real-world networks.


\section{Evaluation} \label{sec:eval}

To illustrate the proposed method for user-objects network we apply it to
Davis's Southern Women dataset \citep{davis1941deep} and the five real-world
datasets presented in section \ref{sec:ubnetworks}. First we are going to
analyse Davis's data in more detail.

\subsection{Southern women}

\begin{figure}[t] \centering
  \begin{tabular}{c} 
  \subfloat[Tf-idf
  weights]{\label{fig:southern_tfidf}\includegraphics[width=14cm]{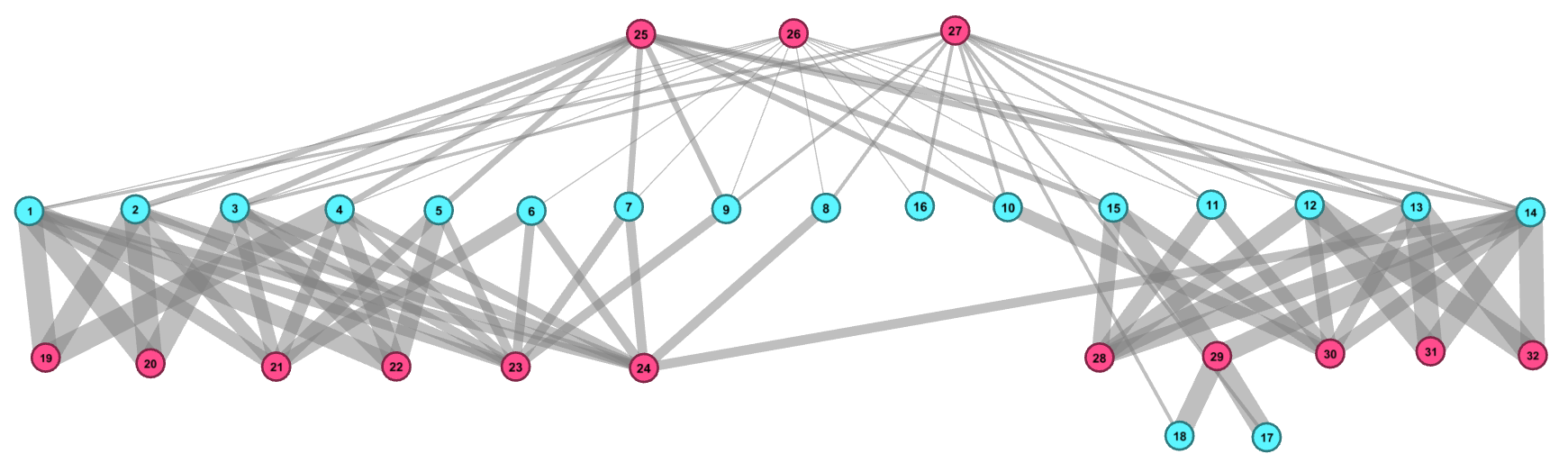}}
  \\   
      \subfloat[Filtered
      network
      $\tau
      =
      1$]{\label{fig:southern_rm}\includegraphics[width=14cm]{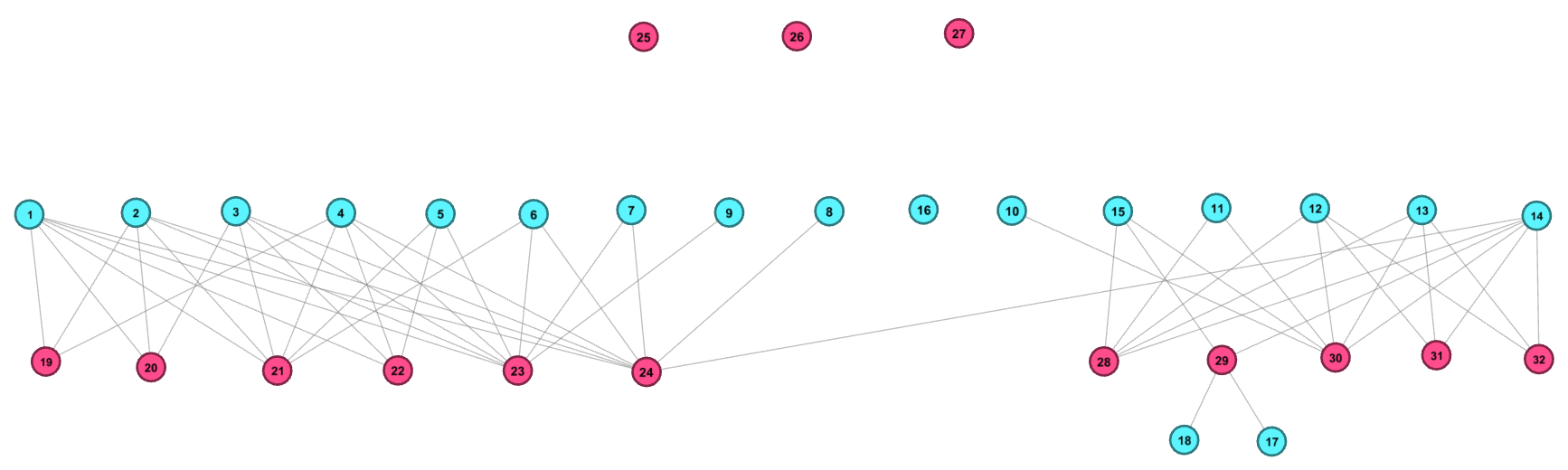}}
  \end{tabular}
  \caption{Southern Women dataset with \emph{tf-idf} weights and filtering
  applied}
  \label{fig:southern}
\end{figure}

The Southern women data was collected by \citet{davis1941deep} during the 1930s
and describes the participation of 18 women in 14 social events. It can be
represented as a bipartite network (figure \ref{fig:southern}) whose nodes are
women (blue nodes) and social events (red nodes). Edges are participation of the
women in the events.

This dataset and the related bipartite network was very well studied by social
scientists. Based on the ethnographic information Davis divided the 18 women
into two groups: women 1-9 in the first group, women 9-18 in the second.
Later, \citet{freeman2003finding} analyzed the outcome of 21 different
studies, which generally identified almost the same two groups, women 1-9 and
10-18. However in some of these studies women 8, 9 and 16 were often identified
as either belonging to both groups or positioned at the periphery of a group.

We applied the proposed weighting scheme to the Southern women dataset and
represented the resulted network in Figure \ref{fig:southern_tfidf}. As in the
previous example, the thickness of an edge is proportional to the \emph{tfidf}
weight. From this figure it is easy to notice that edges connected to very
popular events (25, 26 and 27), where most of women participated, were assigned lower
weights than edges connected to local events (19-24 and 28-32). By applying a
threshold and removing the lower weights, two core groups of women are emerging,
as in Figure \ref{fig:southern_rm}: first group 1-9 and second 10-18. It can be
easily noticed that women 8 and 9 have a weaker connection to the first group
comparing to the other nodes in the group, while woman 16 remains isolated from
both groups.

\subsection{Experiments}

\begin{figure}[t] \centering
  \begin{tabular}{cc} 
  \subfloat[Lastfm]{\label{fig:lastfm_cmp_edges}\includegraphics[width=6.5cm]{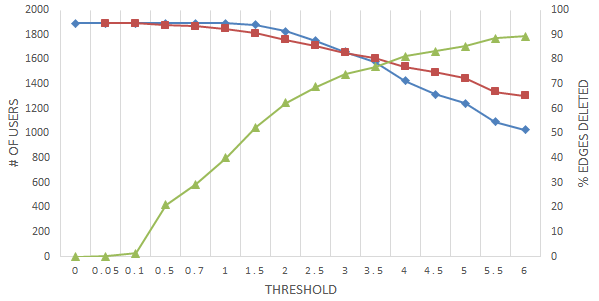}}
  &
  \subfloat[Twitter]{\label{fig:twitter_cmp_edges}\includegraphics[width=6.5cm]{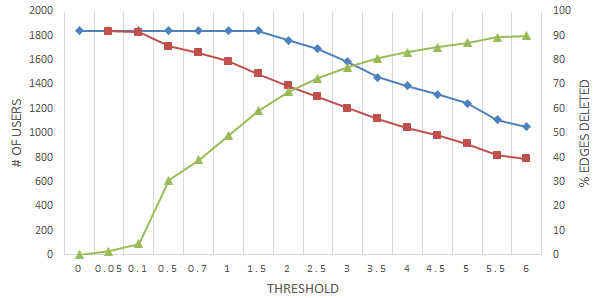}}
  \\
  \subfloat[Audioscrobbler]{\label{fig:audioscrobbler_cmp_edges}\includegraphics[width=6.5cm]{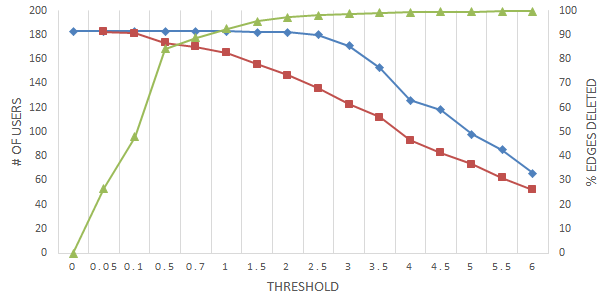}}
  &
  \subfloat[Movielens]{\label{fig:movielens_cmp_edges}\includegraphics[width=6.5cm]{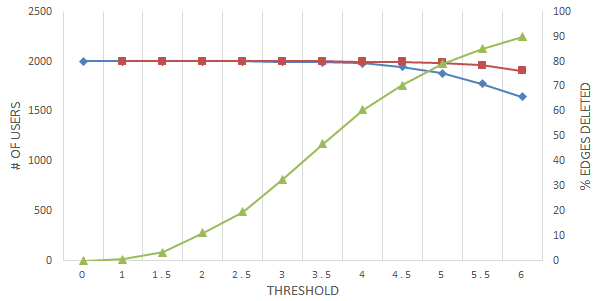}}
  \end{tabular}
  \begin{tabular}{cc} 
  \subfloat[Delicious]{\label{fig:delicious_cmp_edges}\includegraphics[width=6.5cm]{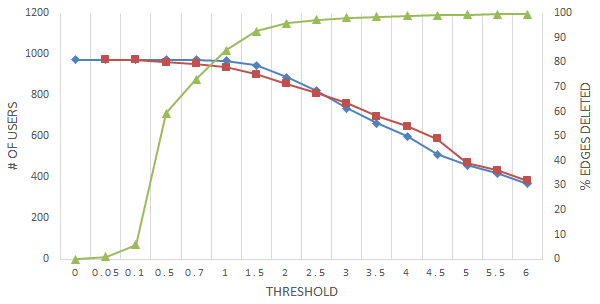}}
  &
  \subfloat{\label{fig:legend_edges}\includegraphics[width=3cm]{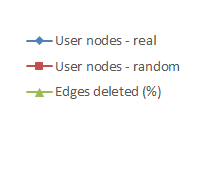}}
  \end{tabular}
  \caption{Ratio of edges (green-triangle) deleted for each threshold and number
  of users in the resulted real (blue-diamond) and random (red-square) networks}
  \label{fig:cmp_edges}
\end{figure}

Next we apply the following method to each of the real world datasets. We
(re)assign the weights of edges based on the \emph{tf-idf} formula as in
equation \ref{formula:weight} and define a range of thresholds $\tau \in \{0.1,
0.5, 1, 1.5, \ldots, 6\}$. For each value $\tau$ a new network $G^{\tau} =
(U^{\tau}, O^{\tau}, E^{\tau})$ is generated by removing the edges with
\emph{tfidf} weights less than the threshold. Remaining nodes with no edges are
removed as well. We project the resulted network on user-nodes ($G^{\tau}_O$)
and apply the Louvain method for community detection \citep{blondel2008fast} in
the projection. The modularity values of the resulted partition will show the
quality of the community structure \citep{newman2004finding}.

In order to demonstrate that the proposed weighting scheme for user-object
networks improves the quality of clustering in the projections while reducing
significantly their density, for each threshold value we generate 100 similar
networks with the same number of edges removed randomly. The resulted random
networks will have the same size (nodes and edges) as the one resulted after
applying the threshold, but instead of deleting edges with \emph{tf-idf} weights
smaller than a threshold we remove them at random. We apply the same community
detection method on projections of each of these networks and compare the
average modularity across all randomly generated networks (random modularity)
with the value resulted from the network where a threshold was applied. If the
random modularity is smaller than the real value then the proposed method
improves quality of the community structure in the projected networks.

In figure \ref{fig:cmp_edges} we represent the percentage of edges removed after
applying each threshold value as above along with the number of users remaining
in both real and random networks. The number of users for random networks was
averaged for all 100 generated instances. It is quite obvious from this figure
that up to a certain threshold value (which depends on the network) the number
of user nodes remains (almost) the same after filtering is applied, even though
a significant number of edges were removed. However, in most randomly generated
networks the number of user nodes decrease at a steeper rate with the threshold.
This observation is particularly visible in Twitter and Audioscrobbler networks
where. after deleting 59\% (threshold 1.5) and 97\% (threshold 2) respectively
of edges, the number of user nodes remains the same. In the meantime, in the
random case in both networks there are around 20\% less user nodes at the same
threshold values. Also, the ratio of edges deleted by applying a threshold is
much higher for Audioscrobbler and Delicious networks comparing to the other
networks. For example, for these two networks 97\%, respectively 96\%, of edges
are removed for a threshold value of 2, while for the other three networks only
11\% (Movielens), 62\% (Lastfm) and 67\% (Twitter) of edges were removed. This
points to a different distribution of \emph{tf-idf} weights for various networks
and this difference has to be considered when selecting the most appropriate
range of thresholds.

By applying the proposed method and filtering out edges up to a threshold value
we are preserving the same number of user nodes (assuming a limit for the
threshold), while significantly reducing the number of edges. The resulted
network will have a lower density and a more clear structure, reflecting users
most important interests.

\begin{figure}[t] \centering
  \begin{tabular}{cc} 
  \subfloat[Lastfm]{\label{fig:lastfm_cmp_density_users}\includegraphics[width=6.5cm]{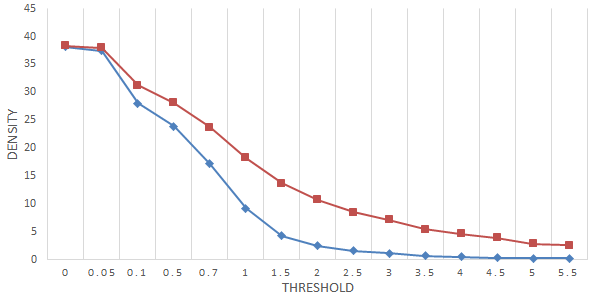}}
  &
  \subfloat[Twitter]{\label{fig:twitter_cmp_density_users}\includegraphics[width=6.5cm]{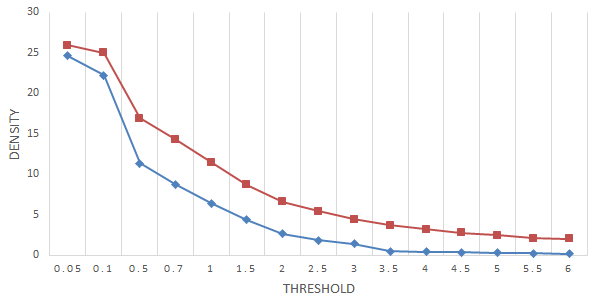}}
  \\
  \subfloat[Audioscrobbler]{\label{fig:audioscrobbler_cmp_density_users}\includegraphics[width=6.5cm]{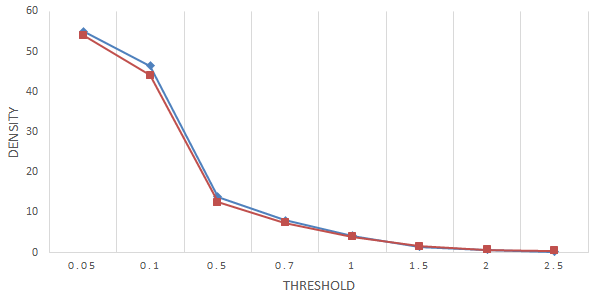}}
  &
  \subfloat[Movielens]{\label{fig:movielens_cmp_density_users}\includegraphics[width=6.5cm]{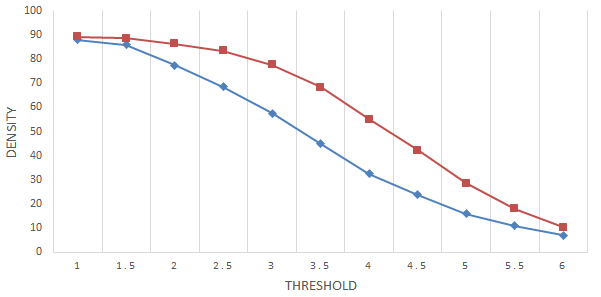}}
  \end{tabular}
  \begin{tabular}{cc} 
  \subfloat[Delicious]{\label{fig:delicious_cmp_density_users}\includegraphics[width=6.5cm]{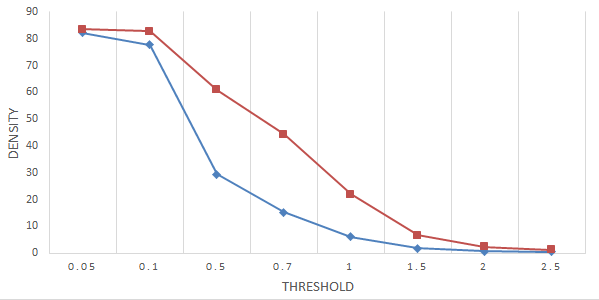}}
  &
  \subfloat{\label{fig:legend_density}\includegraphics[width=3cm]{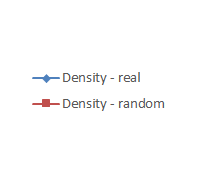}}
  \end{tabular}
  \caption{Density of user-projected network, real (blue-diamond) and random
  (red-square)}
  \label{fig:density_cmp}
\end{figure}

Next we calculate the density of the user-projected networks resulted by
applying the \emph{tfidf} weighting scheme with threshold filtering to the
original network and projecting it on the user nodes (user-projected network).
Both real and random (edges removed randomly) cases are considered for multiple
values of the threshold between 0 and 6. As above, the random density is
averaged for all 100 generated instances. As shown in figure
\ref{fig:density_cmp}, density in real networks is lower than in random
networks, for all datasets. This difference is particularly visible in Lasfm,
Twitter, Movielens and Delicious datasets, where, for some threshold values,
density can be up to 60\% less comparing to random. If we get the Twitter
network for example and select a threshold of 1.5 (no user nodes are removed by
filtering) the density of the user-projected network is 4.3, 82\% less than the
original projected network before filtering (24.6) and 50\% less than the
averaged density for random network (8.67).

\begin{figure}[t] \centering
  \begin{tabular}{cc} 
  \subfloat[Lastfm]{\label{fig:lastfm_cmp_mod_users}\includegraphics[width=6.5cm]{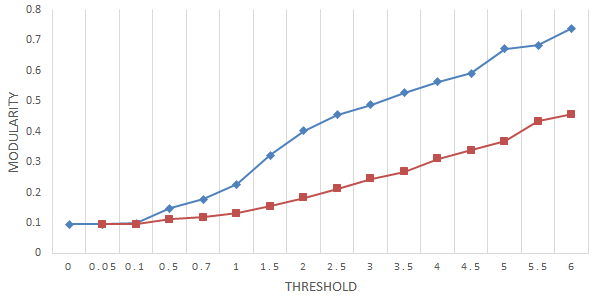}}
  &
  \subfloat[Twitter]{\label{fig:twitter_cmp_mod_users}\includegraphics[width=6.5cm]{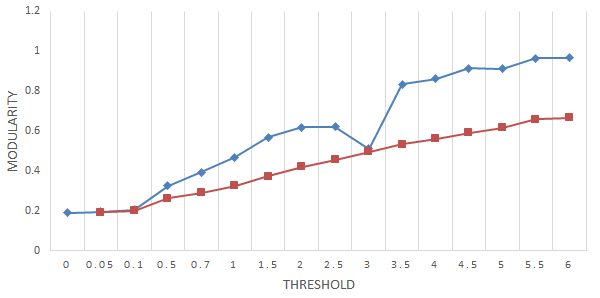}}
  \\
  \subfloat[Audioscrobbler]{\label{fig:audioscrobbler_cmp_mod_users}\includegraphics[width=6.5cm]{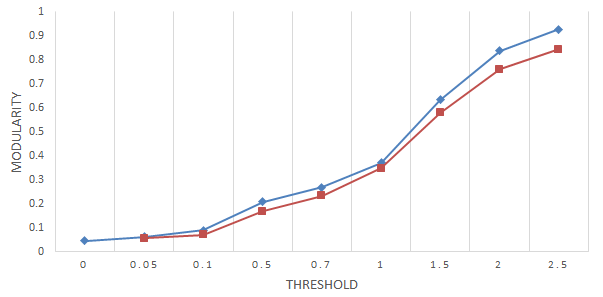}}
  &
  \subfloat[Movielens]{\label{fig:movielens_cmp_mod_users}\includegraphics[width=6.5cm]{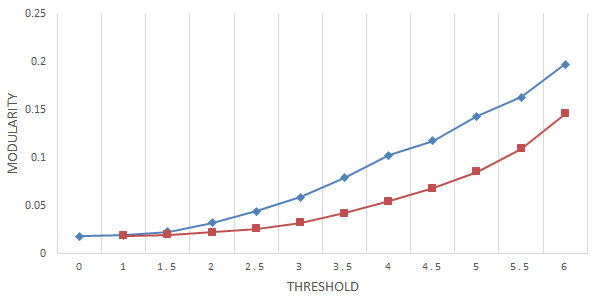}}
  \end{tabular}
  \begin{tabular}{cc} 
  \subfloat[Delicious]{\label{fig:delicious_cmp_mod_users}\includegraphics[width=6.5cm]{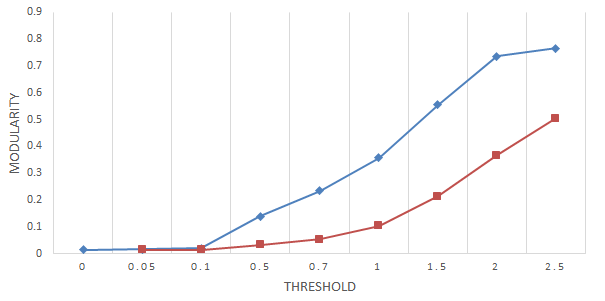}}
  &
  \subfloat{\label{fig:legend_modularity}\includegraphics[width=3cm]{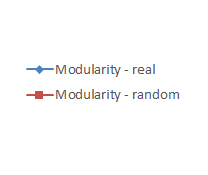}}
  \end{tabular}
  \caption{Modularity values for partitions of user-projected network, real
  (blue-diamond) and random (red-square)}
  \label{fig:mod_cmp}
\end{figure}

Finally, after applying the Louvain method for community detection on the
user-projected networks, we calculate the modularity of the resulted partitions
and compare the values for real and randomly generated networks (the latter
averaged over the 100 random instances). These values, for multiple thresholds,
are represented in figure \ref{fig:mod_cmp}. As mentioned earlier, thresholds
have a bigger impact on Audioscrobbler and Delicious networks with a large part
of edges removed (96-95\%) for a value of 2 or higher. Therefore in these two
cases we limit the threshold to 2.5.

It's easy to notice that modularity is constantly increasing when higher
threshold values are used and more edges are filtered out. Also, for all of our
datasets the modularity values are higher in real networks comparing to random
networks, with this difference being quite significant in most of the cases. For
example, taking the same threshold value of 1.5 for Twitter dataset, the random
modularity is 34\% lower than the modularity for the real projected network.
Similarly, in the Lastfm dataset random modularity is up to 54\% lower (theshold
2) comparing to its value for the real network. The only exception is in Twitter
case for the threshold value of 3 where the two modularities (real and random)
are almost the same. However, we suspect this might be caused by the resolution
limit in modularity and perhaps relative instability of the Louvain method
\citep{fortunato2007resolution}.

From above we saw that applying the proposed \emph{tfidf} method and filtering
out edges below a certain threshold significantly improves both the density
of the bipartite and user-projected networks and the quality of the community
structure of the user-projected network as measured by modularity. This
improvement is also visibly better than filtering out edges randomly without
using the \emph{tfidf} weights. Therefore the resulted network has a simpler and
more intuitive structure, containing only edges that are relevant in finding
groups of similar users.


\section{Conclusion}

We analyzed the bipartite representation of interactions between users and
objects in web or e-commerce systems and shown that due to their scale such
user-object networks have certain characteristics and challenges differentiating
them from other bipartite networks such as collaboration or actor-movie
networks. We analyzed the structural properties of five real world user-object
networks and found a heavy tail degree distribution for objects. This
heterogeneity is responsible for hyperinflation of edges in the projected users
network resulting in very high density and diluted community structure in these
projections. Also, popular objects are connecting together a large number of
users, but they contain little or high level information about users interests.

In order to diminish the impact of higher degree objects we are proposing a new
weighting scheme based on the popular \emph{tf-idf} method used in information
retrieval and text mining. With \emph{tf-idf} the weight of interactions between
users and unpopular objects is amplified while popular objects will bring a
lower contribution to the weight of an edge. Filtering can be applied to keep
only the relevant edges with higher \emph{tf-idf} weights, starting from a
specific threshold. We used the proposed approach with five real world user-object
networks and demonstrate a decrease in density of both original and projected
networks.  We apply the Louvain method \citep{blondel2008fast} to find
communities in the projected users network and calculate the modularity of each
partition. We find that the quality of the community structure as measured by
modularity is significantly improved when compared to the projections of the
original networks. This improvement is also observed with similar partially
random networks where edges are filtered out randomly.
 
We proposed a simple and efficient method of assigning weights in a bipartite
user-object network that will decrease the impact of popular objects and improve
density and community structure. Current work can be extended to develop new
network-based recommender systems. Data used by such systems has a natural
bipartite structure where users are interacting with items in an online
environment. Our approach can also be applied to current community detection
methods for bipartite networks to improve both their scalability and the quality
of results. Further on, due to their simplicity, \emph{tf-idf} weights can be
easily adapted to suit temporal networks that evolve over time. These evolving
weighs can be monitored and analyzed to detect trends and patterns in
user-object event data.

\bibliography{biblio}
\bibliographystyle{year}

\end{document}